\documentclass[prd,nofootinbib,amsfonts,notitlepage,longbibliography]{revtex4-1}
\usepackage[utf8]{inputenc}
\usepackage[T1]{fontenc}
\usepackage{hyperref}
\usepackage{graphicx,bm,amsmath,color,scalerel}
\usepackage{bbold}
\usepackage{wasysym}
\usepackage{verbatim}
\usepackage{amssymb}
\usepackage{epstopdf}
\usepackage{animate}
\usepackage{xmpmulti}
\usepackage{centernot}
\usepackage{multirow}
\usepackage{tikz}
\usepackage{comment}
\usepackage[normalem]{ulem} 
\makeatletter

\newcommand{\be}{\begin{equation}}
\newcommand{\ee}{\end{equation}}
\newcommand{\beq}{\begin{eqnarray}}
\newcommand{\eeq}{\end{eqnarray}}
\newcommand{\ba}{\begin{align}}
\newcommand{\ea}{\end{align}}
\newcommand{\red}[1]{\textcolor{red}{#1}}

\newcommand{\blue}[1]{\textcolor{blue}{#1}}

\begin{document}

\title{Relativistic deformed kinematics from locality conditions in a generalized spacetime}
\author{J.M. Carmona}
\affiliation{Departamento de F\'{\i}sica Te\'orica and Centro de Astropartículas y Física de Altas Energías (CAPA),
Universidad de Zaragoza, Zaragoza 50009, Spain}
\author{J.L. Cort\'es}
\affiliation{Departamento de F\'{\i}sica Te\'orica and Centro de Astropartículas y Física de Altas Energías (CAPA),
Universidad de Zaragoza, Zaragoza 50009, Spain}
\author{J.J. Relancio}
\email{jcarmona@unizar.es, cortes@unizar.es, relancio@unizar.es}
\affiliation{Departamento de F\'{\i}sica Te\'orica and Centro de Astropartículas y Física de Altas Energías (CAPA),
Universidad de Zaragoza, Zaragoza 50009, Spain}

\begin{abstract}
We show how a deformed composition law of four-momenta can be used to define, at the classical level, a modified notion of spacetime for a system of two particles through the crossing of worldlines in particle interactions. We present a derivation of a generic  relativistic isotropic deformed kinematics and discuss the complementarity and relations with other derivations based on $\kappa$-Poincaré Hopf algebra or on the geometry of a maximally symmetric momentum space. 
\end{abstract}

\maketitle

\section{Introduction}


Special relativistic (SR) kinematics is a consequence of the notion of spacetime in Einstein's SR theory. In a quantum theory of gravity (QG), a quantum notion of spacetime will replace the classical notion which leads to SR kinematics. After one hundred years of search for this theory, we still do not have a good testable candidate for QG, partially due to the difficulty in finding observable effects of the theory. This has led to look for alternatives to the purely (unsuccessful) theoretical approaches, opening a recent new approach known as quantum gravity phenomenology~\cite{AmelinoCamelia:2002dx,AmelinoCamelia:2008qg,Mattingly:2005re,Jacobson:2005bg,Liberati2013}. Many works within this new approach are based on the natural expectation that the quantum structure of spacetime will manifest through a modification of the SR kinematics. The consistency with very precise tests of Lorentz invariance~\cite{Colladay:1998fq,Kostelecky:2016pyx,Kostelecky:2016kkn,Kostelecky:2008ts} requires this modification to be parametrized by a new energy scale ($\Lambda$) such that, for observations at energies much smaller than this new scale, the effects of the modification of the SR kinematics are very small. We will refer to this situation as a deformation of SR kinematics (DK). The kinematics of a process (transition between an initial and a final state of free particles) is defined by the expression of the energy of each particle in terms of its momentum (dispersion relation) and by the conservation of the total energy and momentum in the transition, which is determined by the expression of the total energy and momentum of a system of free particles in terms of the energies and momenta of the particles (composition law). A DK will be defined by a deformed dispersion relation (DDR) and a deformed composition law (DCL).               


A possible path to realize the previous ideas is to consider the generalization of Lie algebras as the mathematical framework to implement continuous symmetries in a classical spacetime when one introduces a noncommutativity in spacetime as a first step to the transition to a quantum spacetime. This leads to the formulation of Hopf algebras, whose main new ingredient is a coalgebra structure~\cite{Majid:1995qg}. An example which has played a very  important role in attempts to explore deformations of the SR kinematics is the $\kappa$-Poincaré Hopf algebra~\cite{Majid1994}, which is based on a deformation of the Poincaré Lie algebra and a noncommutative spacetime whose coordinates define a (spatially isotropic) Lie algebra ($\kappa$-Minkowski spacetime). The Casimir of the deformed Poincaré algebra defines a DDR and the coproducts of the translation generators (momentum operators) define a DCL. One can in this way associate a DK to the $\kappa$-Poincaré Hopf algebra~\cite{Lukierski:1991pn}. In fact, the symmetry structure of the Hopf algebra framework translates into a relativistic deformed kinematics (RDK), i.e., a kinematics invariant under new Lorentz transformations connecting different inertial observers. The deformation manifests as a modification of the Lorentz transformations for a one-particle state (determined from the deformation of the Poincaré algebra) and a (nontrivial) modification of the Lorentz transformation of a two-particle system (determined from the nontrivial coproduct of the Lorentz generators).

The idea to consider a relativistic theory with a second invariant (a length $\ell$), on top of the velocity $c$, was motivated by the appearance of a minimal length~\cite{Garay1995,Hossenfelder:2012jw} in different approaches to QG. This led to consider a DDR with a new scale appearing as a cutoff on the energy or on the momentum~\cite{Bruno2001}  as examples of a doubly special relativity (DSR). The nonlinearity of the Lorentz transformations which leave the DDR invariant implies a nonlinearity of the composition law which should be determined by the invariance under Lorentz transformations of the conservation of the total energy and momentum defined by the nonlinear composition law. The study of these examples led to identify their relation with the $\kappa$-Poincaré kinematics determined in the Hopf algebra framework~\cite{KowalskiGlikman:2002we}.

More recently, a new approach to a deformation of SR kinematics was introduced based on a model for the interaction of particles defined by a DCL. The crossing of worldlines which characterizes the interaction of particles in the case (SR) of a linear composition law does no longer happen due to the deformation of the composition law. Locality of the interaction (for any observer) is lost; only the observer whose origin is on the interaction sees a crossing of worldlines. The locality of interactions in SR is replaced by a relative locality (RL)~\cite{AmelinoCamelia:2011bm}. The DCL could be used to define a connection in momentum space and, together with the identification of a DDR from the distance between the origin and a point in momentum space, one has an interpretation of a DK based on the geometry of momentum space~\cite{Gubitosi:2013rna}.

In Ref.~\cite{Carmona:2019fwf}, it has been shown that if one considers a maximally symmetric momentum space and chooses coordinates in momentum space such that the metric is spatially isotropic, one can define a DDR from the distance between the origin and a point in momentum space calculated with the metric, and also a DCL from the isometries of the metric which do not leave the origin invariant (translations in momentum space). One can show that the deformed kinematics defined by the metric is a RDK. This gives an alternative simple relation between a deformed kinematics and a geometry in momentum space (the scale of deformation is related to the curvature in momentum space). It also allows (in contrast with the relation between the geometry of momentum space and the kinematics based on relative locality) to implement the relativistic invariance in a simple way. The Lorentz invariance of the DDR is a direct consequence of the identification of the Lorentz transformation of a one-particle state with the isometries of the metric which leave the origin invariant. The Lorentz invariance of the conservation law defined by the DCL can be understood within the geometric framework through the identification of the DCL and the Lorentz transformations as isometries~\cite{Carmona:2019fwf}.

We are going to follow in this work a different path in the study of a DK. The idea is to take the model for the interaction of particles defined by a DCL and try to go from the loss of locality in the spacetime whose coordinates are the canonical coordinates of a phase space together with the four-momentum coordinates, to a new set of space-time coordinates in phase space such that all the particles have the same coordinates at the interaction. The interaction defined by a DCL is then local in a generalized two-particle spacetime defined as a nontrivial subspace of the two-particle phase space. In Ref.~\cite{Carmona2018}, an ansatz for the new space-time coordinates of each of the particles, defined as a linear combination of their space-time coordinates with coefficients depending on the momenta of both particles, was introduced. The locality of the interaction in the new spacetime leads to a system of differential equations relating the functions of momenta which define the new space-time coordinates and the DCL. When one assumes that the new space-time coordinates of one of the particles do not depend on the four-momentum of the other particle\footnote{This can only be the case for one of the particles, otherwise the new space-time coordinates are commutative and then one can always find a change of momentum variables which leads to SR kinematics~\cite{Carmona2018}.} and that they are just a representation of $\kappa$-Minkowski noncommutative spacetime, then the equations derived from the locality of the interaction can be used to determine the DCL. If one uses the representation of $\kappa$-Minkowski which reproduces the phase-space structure of the $\kappa$-Poincaré Hopf algebra in the bicrossproduct basis, the corresponding DCL determined by locality turns out to be the one corresponding to the $\kappa$-Poincaré kinematics. This result shows that $\kappa$-Poincaré relativistic kinematics can be seen as an example of a deformed kinematics compatible with the possibility to identify a new spacetime where interactions are local. In Ref.~\cite{Kowalski-Glikman:2014wba}, one arrives to the same conclusion from a related perspective. One takes an extension to $3+1$ dimensions of the model for the interaction of particles in $2+1$ dimensions and finds that implementing the rigidity of translations (which is one way to rephrase the requirement of locality of interactions), one reproduces the $\kappa$-Poincaré relativistic kinematics.      

In the present paper we go a step further in the relationship between a relativistic deformed kinematics and the definition of generalized space-time coordinates where interactions are local. While Ref.~\cite{Carmona2018} did not include any restriction on the DCL and the noncommutative spacetime that implements locality, here we will consider a different ansatz for implementing locality:\footnote{An ansatz is required in order to give a content to the requirement of locality of interactions. Otherwise it would always be possible to find a change of variables in the two-particle phase space such that the composition law reduces to the sum of four-momenta. In this case one would be just considering SR in a fancy choice of coordinates.} the new space-time coordinates of each particle are linear combinations of the space-time coordinates of both particles, but the coefficients of the space-time coordinates of each particle depend only on its momentum. This definition of new space-time coordinates in the phase space of the two-particle system may be seen as a more natural prescription than the one made in Ref.~\cite{Carmona2018}, where the generalized space-time coordinates depend on both momenta but do not mix the space-time coordinates of the two particles. Moreover, the new ansatz imposes a strong condition on the DCL, so that not every composition law can lead to local interactions.  

The structure of the paper goes as follows. In Sec.~\ref{sec_st_locality}, we define a generalized two-particle system spacetime which implements the locality of interactions with a DCL. In the new space-time coordinates, one has a sum of two contributions, each one involving the phase-space coordinates of one of the particles. As we show in Sec.~\ref{sec_first_order}, the new system of equations relating the functions of one four-momentum which define the new space-time coordinates and the derivatives of the DCL can in this case be used to directly determine the DCL when one makes the ansatz that the deformed composition law contains only terms proportional to the inverse of the scale of deformation $\Lambda$ (DCL1). We will later show that such a locality-compatible DCL1 corresponds to the $\kappa$-Poincaré composition law in a basis that is different from the bicrossproduct basis (which is the most widely used in $\kappa$-Poincaré studies). Once we have obtained the composition law, we study the noncommutativity of the one-particle and two-particle spacetimes defined by the locality of interactions. 

In Sec.~\ref{sec_rel_kinematics}, we determine the corresponding DDR which, together with a DCL1 compatible with locality, defines a RDK, and the nonlinear Lorentz transformations of the one-particle and two-particle systems. This provides a new way to derive a RDK based on the physical principle of locality of interactions, which is an alternative to the more formal derivations of a RDK based on $\kappa$-Poincaré Hopf algebra or on the geometry of a maximally symmetric momentum space.  
As we will see in Sec.\ref{sec_comparison}, the new derivation of the relativistic deformed kinematics based on locality (or on the geometry of a maximally symmetric moment space) not only reproduces the results based on the $\kappa$-Poincaré Hopf algebra but also identifies an alternative in which the new energy scale of the deformation does not appear as a maximum energy.
Then, in Sec.~\ref{sec:associativity}, we will study the role of associativity in the definition of a RDK and will conclude that an associative DCL1 (which corresponds to $\kappa$-Poincaré in a certain basis), which is locality-compatible, is the only relativistic isotropic generalization of SR kinematics.

We end up in Sec.~\ref{sec:conclusions} with a summary and prospects for further work.
 
\section{Spacetime from locality}
\label{sec_st_locality}


We consider the classical model for the interaction of two particles with a deformed kinematics defined by the action
\begin{align}
S \,=& \int_{-\infty}^0 d\tau \sum_{i=1,2} \left[x_{-(i)}^\mu(\tau) \dot{p}_\mu^{-(i)}(\tau) + N_{-(i)}(\tau) \left[C(p^{-(i)}(\tau)) - m_{-(i)}^2\right]\right] \nonumber \\   
& + \int^{\infty}_0 d\tau \sum_{j=1,2} \left[x_{+(j)}^\mu(\tau) \dot{p}_\mu^{+(j)}(\tau) + N_{+(j)}(\tau) \left[C(p^{+(j)}(\tau)) - m_{+(j)}^2\right]\right] \nonumber \\
& + \xi^\mu \left[{\cal P}^+_\mu(0) - {\cal P}^-_\mu(0)\right]\,,
\label{S1}
\end{align}
where $\dot{a}\doteq (da/d\tau)$ is a derivative of the variable $a$ with respect to the parameter $\tau$ along the trajectory of the particle, $x_{-(i)}$ ($x_{+(j)}$) are the space-time coordinates of the in-state (out-state) particles, $p^{-(i)}$ ($p^{+(j)}$) their four-momenta, $m_{-(i)}$ ($m_{+(j)}$) their masses, ${\cal P}^-$ (${\cal P}^+$) the total four-momentum of the in-state (out-state) defining the DCL, $C(k)$ the function of a four-momentum $k$ defining the DDR, $\xi^\mu$ are Lagrange multipliers that implement the energy-momentum conservation in the interaction, and $N_{-(i)}$ ($N_{+(j)}$) are Lagrange multipliers implementing the dispersion relation of in-state (out-state) particles.

The variational principle applied to the action (\ref{S1})  fixes the end (starting) space-time coordinates of the trajectories of the in-state (out-state) particles
\be
x_{-(i)}^\mu(0) \,=\, \xi^\nu \frac{\partial {\cal P}^-_\nu}{\partial p^{-(i)}_\mu}(0)\,, \quad\quad\quad
x_{+(j)}^\mu(0) \,=\, \xi^\nu \frac{\partial {\cal P}^+_\nu}{\partial p^{+(j)}_\mu}(0)\,.
\ee


When the total four-momentum is just the sum of the four-momenta of the particles, one has $x_{-(i)}^\mu(0)=x_{+(j)}^\mu(0)=\xi^\mu$ and the worldlines of the four particles cross at the point with coordinates $\xi^\mu$ (local interaction). When one has a DCL, the locality of the interaction is lost. 

We now ask the question whether it is possible to identify new space-time coordinates in the phase space of the two particles (we consider either the two particles in the in-state or out-state and then omit the index $-$,$+$),
\be
\tilde{x}^\alpha_{(1)} \,=\, x^\mu_{(1)} \varphi^\alpha_\mu(p^{(1)}) + x^\mu_{(2)} \varphi^{(2)\alpha}_{(1)\mu}(p^{(2)})\,, \quad\quad
\tilde{x}^\alpha_{(2)} \,=\, x^\mu_{(2)} \varphi^\alpha_\mu(p^{(2)}) + x^\mu_{(1)} \varphi^{(1)\alpha}_{(2)\mu}(p^{(1)})\,,
\ee
such that the interaction is local in the new spacetime ($\tilde{x}^\alpha_{(1)}(0)=\tilde{x}^\alpha_{(2)}(0)$). We assume that $\varphi^{(2)\alpha}_{(1)\mu}(0) = \varphi^{(1)\alpha}_{(2)\mu}(0) = 0$ so that the system of two particles reduces, when one of the two momenta is zero, to one particle with new space-time coordinates $\tilde{x}^\alpha = x^\mu \varphi^\alpha_\mu(p)$. One also has $\varphi^\alpha_\mu(0)=\delta^\alpha_\mu$ so that the new space-time coordinates coincide with the coordinates $x$ in the limit $p\to 0$.  

Locality in the generalized spacetime requires to find a set of functions $\varphi^\alpha_\mu(k)$, $\varphi^{(2)\alpha}_{(1)\mu}(k)$ and $\varphi^{(1)\alpha}_{(2)\mu}(k)$ satisfying the set of equations
\be
\frac{\partial(p^{(1)}\oplus p^{(2)})_\mu}{\partial p^{(1)}_\nu} \,\varphi^\alpha_\nu(p^{(1)}) \,+\, 
\frac{\partial(p^{(1)}\oplus p^{(2)})_\mu}{\partial p^{(2)}_\nu} \,\varphi^{(2)\alpha}_{(1)\nu}(p^{(2)}) \,=\, \frac{\partial(p^{(1)}\oplus p^{(2)})_\mu}{\partial p^{(2)}_\nu} \,\varphi^\alpha_\nu(p^{(2)}) \,+\, 
\frac{\partial(p^{(1)}\oplus p^{(2)})_\mu}{\partial p^{(1)}_\nu} \,\varphi^{(1)\alpha}_{(2)\nu}(p^{(1)})\,,
\label{loc-eq}
\ee
where we use the notation 
\be
{\cal P}_\mu \,=\, (p^{(1)}\oplus p^{(2)})_\mu\,,
\ee
for the components of the total four-momentum (${\cal P}$) of a system of two particles with four-momentum $p^{(1)}$, $p^{(2)}$. We will refer to $\oplus$ as the DCL. 

Eqs. (\ref{loc-eq}) are just the condition that the worldlines of the two particles in the in-state
(or in the out-state) cross at a point. But the four-momentum of the two particles in the in-state and out-state are constrained by the conservation of the total four-momentum
\be
(p^{-(1)}\oplus p^{-(2)})_\mu \,=\, (p^{+(1)}\oplus p^{+(2)})_\mu\,.
\ee
Then, the crossing of the worldlines of the four particles at a point requires the left-hand side and the right-hand side of Eqs. (\ref{loc-eq}) to depend on the two four-momenta only through the combination $(p^{(1)}\oplus p^{(2)})$. When one uses the conditions  $\varphi^{(2)\alpha}_{(1)\mu}(0) = \varphi^{(1)\alpha}_{(2)\mu}(0) = 0$, one concludes that in fact both sides of Eq.~(\ref{loc-eq}) should be equal to $\varphi^\alpha_\mu(p^{(1)}\oplus p^{(2)})$.\footnote{A way to see this is to consider the situation in which the particles in the in-state have four-momentum $p_\mu^{-(1)}=(p^{+(1)}\oplus p^{+(2)})_\mu$, $p_\mu^{-(2)}=0$.}

When one takes the limit $p^{(1)}\to 0$ or $p^{(2)}\to 0$ in the locality equations, one has
\be
\varphi^{(2)\alpha}_{(1)\mu}(p^{(2)}) \,=\, \varphi^\alpha_\mu(p^{(2)}) - \lim_{k\to 0}  \frac{\partial(k\oplus p^{(2)})_\mu}{\partial k_\alpha}\,, \quad\quad
\varphi^{(1)\alpha}_{(2)\mu}(p^{(1)}) \,=\, \varphi^\alpha_\mu(p^{(1)}) - \lim_{k\to 0} \frac{\partial(p^{(1)}\oplus k)_\mu}{\partial k_\alpha}\,,
\label{eq:phi12}
\ee
for the functions which define the mixing of phase spaces of the two particles in the generalized space-time coordinates. 


When these expressions for the mixing functions $\varphi^{(2)}_{(1)}$, $\varphi^{(1)}_{(2)}$ are plugged into the locality equations, one finds
\begin{align}
& \frac{\partial(p^{(1)}\oplus p^{(2)})_\mu}{\partial p^{(2)}_\nu} \, \lim_{k\to 0} \frac{\partial(k\oplus p^{(2)})_\nu}{\partial k_\alpha} \,=\, \frac{\partial(p^{(1)}\oplus p^{(2)})_\mu}{\partial p^{(1)}_\nu}  \, \lim_{k\to 0} \frac{\partial(p^{(1)}\oplus k)_\nu}{\partial k_\alpha} \,=\, \nonumber \\ & \:\:\: \frac{\partial(p^{(1)}\oplus p^{(2)})_\mu}{\partial p^{(1)}_\nu} \varphi^\alpha_\nu(p^{(1)}) + 
\frac{\partial(p^{(1)}\oplus p^{(2)})_\mu}{\partial p^{(2)}_\nu} \varphi^\alpha_\nu(p^{(2)}) - \varphi^\alpha_\mu(p^{(1)}\oplus p^{(2)})\,.
\label{loc-oplus-varphi}
\end{align}
The first equality is a set of equations that a DCL ($\oplus$) has to satisfy in order to be able to have a generalized spacetime (whose coordinates are a sum of two terms, each one involving the phase space coordinates of a particle) where interactions are local. The second equality is a set of relations between the DCL ($\oplus$) and the functions $\varphi^\alpha_\mu$ which define the new space-time coordinates for a one-particle system. 


We introduce the relative coordinate
\begin{align}
\tilde{x}^\alpha_{(12)} \,\doteq\, \tilde{x}_{(1)}^\alpha - \tilde{x}_{(2)}^\alpha & = x_{(1)}^\mu \left[\varphi^\alpha_\mu(p^{(1)}) - \varphi^{(2)\alpha}_{(1)\mu}(p^{(1)})\right] - x_{(2)}^\mu 
\left[\varphi^\alpha_\mu(p^{(2)}) - \varphi^{(1)\alpha}_{(2)\mu}(p^{(2)})\right] \nonumber \\
&= x_{(1)}^\mu \,\lim_{k\to 0} \frac{\partial(p^{(1)}\oplus k)_\mu}{\partial k_\alpha} - x_{(2)}^\mu \,\lim_{k\to 0} \frac{\partial(k\oplus p^{(2)})_\mu}{\partial k_\alpha}\,.
\label{xtilde}
\end{align}
The effect of an infinitesimal transformation with parameters $\epsilon^\mu$ generated by the total four-momentum (translation) on the relative coordinate is
\be
\delta\tilde{x}^\alpha_{(12)} \,=\, \epsilon^\mu \{\tilde{x}^\alpha_{(12)}, (p^{(1)}\oplus p^{(2)})_\mu\} \,=\, \epsilon^\mu \left[- \frac{\partial(p^{(1)}\oplus p^{(2)})_\mu}{\partial p^{(1)}_\nu} \lim_{k\to 0} \frac{\partial(p^{(1)}\oplus k)_\nu}{\partial k_\alpha} + \frac{\partial(p^{(1)}\oplus p^{(2)})_\mu}{\partial p^{(2)}_\nu} \lim_{k\to 0} \frac{\partial(k\oplus p^{(2)})_\nu}{\partial k_\alpha}\right]\,.
\ee 
We see then that the system of equations that a DCL ($\oplus$) has to satisfy in order to find a generalized space-time coordinates with a local interaction is just the condition of invariance of the relative coordinate under translations. If one observer sees a crossing of worldlines ($\tilde{x}^\alpha_{(12)}(0)=0$), another observer related by a translation also sees a crossing of worldlines. 


One can also consider the effect of an infinitesimal transformation with parameters $\epsilon_\alpha$ generated by the relative generalized space-time coordinates $\tilde{x}^\alpha_{(12)}$ on the momenta $p^{(1)}$, $p^{(2)}$. One has 
\begin{align}
\delta p^{(1)}_\mu & = \epsilon_\alpha \{p^{(1)}_\mu, \tilde{x}^\alpha_{(12)}\} \,=\, \epsilon_\alpha \lim_{k\to 0} \frac{\partial(p^{(1)}\oplus k)_\mu}{\partial k_\alpha} \,=\, \left[(p^{(1)}\oplus \epsilon) - p^{(1)}\right]_\mu\,, \nonumber \\
\delta p^{(2)}_\mu & = \epsilon_\alpha \{p^{(2)}_\mu, \tilde{x}^\alpha_{(12)}\} \,=\, -\epsilon_\alpha \lim_{k\to 0} \frac{\partial(k\oplus p^{(2)})_\mu}{\partial k_\alpha} \,=\, -\left[(\epsilon\oplus p^{(2)}) - p^{(2)}\right]_\mu\,,
\end{align}
and then 
\be
\delta(p^{(1)}\oplus p^{(2)}) \,=\, \delta p^{(1)} \oplus p^{(2)} + p^{(1)}\oplus \delta p^{(2)} \,=\, (p^{(1)}\oplus \epsilon)\oplus p^{(2)} - p^{(1)}\oplus(\epsilon\oplus p^{(2)})\,.
\label{delta_p1+p2}
\ee
But the invariance of the relative coordinate under the transformation generated by the total four-momentum implies the invariance of the total four-momentum under the transformation generated by the relative coordinate. Then, from Eq.~\eqref{delta_p1+p2}, one has 
\be
(p^{(1)}\oplus \epsilon)\oplus p^{(2)} \,=\, p^{(1)}\oplus(\epsilon\oplus p^{(2)})\,.
\label{eq:associativity}
\ee
An alternative more direct way to derive this result is based on the identities
\begin{align}
& \frac{\partial(p^{(1)}\oplus p^{(2)})_\mu}{\partial p^{(2)}_\nu} \, \lim_{k\to 0} \frac{\partial(k\oplus p^{(2)})_\nu}{\partial k_\alpha} \,=\, \lim_{k\to 0}  \frac{\partial(p^{(1)}\oplus (k\oplus p^{(2)}))_\mu}{\partial (k\oplus p^{(2)})_\nu} \, \frac{\partial(k\oplus p^{(2)})_\nu}{\partial k_\alpha} \,=\, \lim_{k\to 0} \frac{\partial(p^{(1)}\oplus (k\oplus p^{(2)}))_\mu}{\partial k_\alpha}\,, \\ \nonumber
& \frac{\partial(p^{(1)}\oplus p^{(2)})_\mu}{\partial p^{(1)}_\nu}  \, \lim_{k\to 0} \frac{\partial(p^{(1)}\oplus k)_\nu}{\partial k_\alpha} \,=\,  \lim_{k\to 0} \,\frac{\partial((p^{(1)} \oplus k)\oplus p^{(2)})_\mu}{\partial (p^{(1)}\oplus k)_\nu}  \, \frac{\partial(p^{(1)}\oplus k)_\nu}{\partial k_\alpha} \,=\, \lim_{k\to 0} \frac{\partial((p^{(1)}\oplus k)\oplus p^{(2)})_\mu}{\partial k_\alpha}\,.
\end{align}
Then, the first equality in (\ref{loc-oplus-varphi}) leads to 
\be
\lim_{k\to 0} \frac{\partial(p^{(1)}\oplus (k\oplus p^{(2)}))_\mu}{\partial k_\alpha} \,=\, \lim_{k\to 0} \frac{\partial((p^{(1)}\oplus k)\oplus p^{(2)})_\mu}{\partial k_\alpha}\,,
\ee
which is equivalent to Eq.~\eqref{eq:associativity}.

If one makes the choice $\varphi^{(2)\alpha}_{(1)\mu}(p^{(2)})=0$ in Eq.~\eqref{eq:phi12},\footnote{The same argument can be made if one takes the alternative choice $\varphi^{(1)\alpha}_{(2)\mu}(p^{(1)})=0$.} one has
\begin{equation}
\begin{split}
\frac{\partial(p^{(1)}\oplus p^{(2)})_\mu}{\partial p^{(1)}_\nu} \,\varphi^\alpha_\nu(p^{(1)}) \,=\, \frac{\partial(p^{(1)}\oplus p^{(2)})_\mu}{\partial p^{(1)}_\nu} \,\lim_{k\to 0} \frac{\partial(k\oplus p^{(1)})_\nu}{\partial k_\alpha} \,&=\, \lim_{k\to 0} \left[\frac{\partial((k\oplus p^{(1)})\oplus p^{(2)})_\mu}{\partial(k\oplus p^{(1)})_\nu} \,\frac{\partial(k\oplus p^{(1)})_\nu}{\partial k_\alpha}\right]\\
\,&=\, \lim_{k\to 0} \frac{\partial((k\oplus p^{(1)})\oplus p^{(2)})_\mu}{\partial k_\alpha}\,,  \\
\frac{\partial(p^{(1)}\oplus p^{(2)})_\mu}{\partial p^{(2)}_\nu} \,\varphi^\alpha_\nu(p^{(2)}) \,=\, \frac{\partial(p^{(1)}\oplus p^{(2)})_\mu}{\partial p^{(2)}_\nu} \,\lim_{k\to 0} \frac{\partial(k\oplus p^{(2)})_\nu}{\partial k_\alpha} \,&=\, \lim_{k\to 0} \left[\frac{\partial(p^{(1)}\oplus(k\oplus p^{(2)})_\mu}{\partial(k\oplus p^{(2)})_\nu} \,\frac{\partial(k\oplus p^{(2)})_\nu}{\partial k_\alpha}\right]\\ \,&=\, \lim_{k\to 0} \frac{\partial(p^{(1)}\oplus(k\oplus p^{(2)})_\mu}{\partial k_\alpha}\,, \\
\varphi^\alpha_\mu(p^{(1)}\oplus p^{(2)}) \,&=\, \lim_{k\to 0} \frac{\partial(k\oplus(p^{(1)}\oplus p^{(2)}))_\mu}{\partial k_\alpha}\,.
\end{split}
\end{equation}

Then the relations of compatibility with locality, Eqs.~\eqref{loc-oplus-varphi}, can be written as
\be
\begin{split}
\lim_{k\to 0} \frac{\partial(p^{(1)}\oplus (k\oplus p^{(2)}))_\mu}{\partial k_\alpha} \,&=\, \lim_{k\to 0} \frac{\partial((p^{(1)}\oplus k)\oplus p^{(2)})_\mu}{\partial k_\alpha} \\ \,&=\, \lim_{k\to 0} \frac{\partial((k\oplus p^{(1)})\oplus p^{(2)})_\mu}{\partial k_\alpha} \,+\, \lim_{k\to 0} \frac{\partial(p^{(1)}\oplus(k\oplus p^{(2)}))_\mu}{\partial k_\alpha} \,-\,\lim_{k\to 0} \frac{\partial(k\oplus(p^{(1)}\oplus p^{(2)}))_\mu}{\partial k_\alpha}\,.
\end{split}
\ee
This makes manifest that any associative DCL is compatible with locality.

\section{First-order deformed composition law of four-momenta (DCL1)}
\label{sec_first_order}

We consider a DCL which is linear as a function of the four-momentum of each particle. Dimensional arguments lead to the general form for such deformed composition law (DCL1)
\be
(p^{(1)}\oplus p^{(2)})_\mu \,=\, p^{(1)}_\mu + p^{(2)}_\mu + \frac{c_\mu^{\nu\rho}}{\Lambda} p^{(1)}_\nu p^{(2)}_\rho\,,
\ee
where $c_\mu^{\nu\rho}$ are arbitrary dimensionless coefficients. Let us see if such DCL can satisfy the restrictions from locality of interactions. One has 
\begin{align}
& \frac{\partial(p^{(1)}\oplus p^{(2)})_\mu}{\partial p^{(2)}_\nu} \,=\, \delta^\nu_\mu + \frac{c_\mu^{\rho\nu}}{\Lambda} p^{(1)}_\rho\,, \quad\quad \lim_{k\to 0} \frac{\partial(k\oplus p^{(2)})_\nu}{\partial k_\alpha} \,=\, \delta^\alpha_\nu + \frac{c^{\alpha\sigma}_\nu}{\Lambda} p^{(2)}_\sigma\,, \\ \nonumber 
& \frac{\partial(p^{(1)}\oplus p^{(2)})_\mu}{\partial p^{(1)}_\nu} \,=\, \delta^\nu_\mu + \frac{c_\mu^{\nu\sigma}}{\Lambda} p^{(2)}_\sigma\,, \quad\quad  \lim_{k\to 0} \frac{\partial(p^{(1)}\oplus k)_\nu}{\partial k_\alpha} \,=\, \delta^\alpha_\nu + \frac{c_\nu^{\rho\alpha}}{\Lambda} p^{(1)}_\rho\,,
\end{align}
and
\begin{align}
& \frac{\partial(p^{(1)}\oplus p^{(2)})_\mu}{\partial p^{(2)}_\nu} \, \lim_{k\to 0} \frac{\partial(k\oplus p^{(2)})_\nu}{\partial k_\alpha} \,=\, \delta^\alpha_\mu + \frac{c_\mu^{\rho\alpha}}{\Lambda} p^{(1)}_\rho + \frac{c^{\alpha\sigma}_\mu}{\Lambda} p^{(2)}_\sigma + \frac{c_\mu^{\rho\nu} c^{\alpha\sigma}_\nu}{\Lambda^2} p^{(1)}_\rho p^{(2)}_\sigma\,, \nonumber \\
& \frac{\partial(p^{(1)}\oplus p^{(2)})_\mu}{\partial p^{(1)}_\nu}  \, \lim_{k\to 0} \frac{\partial(p^{(1)}\oplus k)_\nu}{\partial k_\alpha} \,=\,  \delta^\alpha_\mu + \frac{c_\mu^{\rho\alpha}}{\Lambda} p^{(1)}_\rho + \frac{c^{\alpha\sigma}_\mu}{\Lambda} p^{(2)}_\sigma + \frac{c_\mu^{\nu\sigma} c^{\rho\alpha}_\nu}{\Lambda^2} p^{(1)}_\rho p^{(2)}_\sigma\,.
\end{align}  
A DCL1 is compatible with locality if the dimensionless coefficients satisfy the system of equations 
\be
c_\mu^{\rho\nu} c^{\alpha\sigma}_\nu \,=\, c_\mu^{\nu\sigma} c^{\rho\alpha}_\nu\,.
\ee
These are just the conditions that the coefficients $c_\mu^{\nu\rho}$ of a DCL1 have to satisfy in order to be associative. This result can be understood since Eq.~\eqref{eq:associativity} implies associativity for a DCL1.


The general form of an isotropic DCL1 has coefficients 
\be
c_\mu^{\nu\rho} \,=\, c_1 \,\delta_\mu^\nu n^\rho + c_2 \,\delta_\mu^\rho n^\nu + c_3 \,\eta^{\nu\rho} n_\mu + c_4 \,n_\mu n^\nu n^\rho + c_5 \,\epsilon_\mu^{\:\:\nu\rho\sigma} n_\sigma\,,
\ee
where $n_\mu = (1, 0, 0, 0)$ and $c_i$ are arbitrary constants. Compatibility with locality leads to four possible cases for the DCL1:
\be
c_\mu^{\nu\rho} \,=\, \delta_\mu^\rho n^\nu\,, \quad\quad\quad  c_\mu^{\nu\rho} \,=\, \delta_\mu^\nu n^\rho\,, \quad\quad\quad c_\mu^{\nu\rho} \,=\, \delta_\mu^\nu n^\rho + \delta_\mu^\rho n^\nu - n_\mu n^\nu n^\rho\,, \quad\quad\quad  c_\mu^{\nu\rho} \,=\, \eta^{\nu\rho} n_\mu - n_\mu n^\nu n^\rho\,.
\ee
In the last two cases, corresponding to a symmetric composition law, it is possible to find a change of the choice of four-momentum variables ($k'_\mu = f_\mu(k)$) such that the composition in the new variables reduces to the addition of momenta ($(p'\oplus' q')_\mu \doteq (p\oplus q)'_\mu=p'_\mu + q'_\mu$)\footnote{For the first of them, the function is $f_0(k)=\Lambda \log(1+k_0/\Lambda)$, $f_i(k)=k_i/(1+k_0/\Lambda)$, while for the last one $f_0(k)=k_0-\vec{k}^2/(2\Lambda)$, $f_i(k)=k_i$.}. Then, they do not correspond to a deformation of SR based on a deformed composition law.

In the remaining two cases, one has a non-symmetric composition law (in fact the two cases are related by an exchange of the four-momenta in the composition law). A change of four-momentum variables applied to an additive composition law will always produce a symmetric composition law; therefore, the two cases of a non-symmetric composition law are real deformations of SR. The explicit form of the locality-compatible DCL1 (or, for short, ``local'' DCL1) is\footnote{There is another DCL1 obtained by exchanging the four-momentum variables.}
\be
(p^{(1)}\oplus p^{(2)})_0 \,=\, p^{(1)}_0 + p^{(2)}_0 + \epsilon\, \frac{p^{(1)}_0 p^{(2)}_0}{\Lambda}\,, \quad\quad\quad\quad
(p^{(1)}\oplus p^{(2)})_i \,=\, p^{(1)}_i + p^{(2)}_i + \epsilon\, \frac{p^{(1)}_0 p^{(2)}_i}{\Lambda}\,,
\label{DCL(1)}
\ee
where $\epsilon = \pm 1$ is an overall sign for the modification in the composition law and an arbitrary constant can be reabsorbed in the definition of the scale $\Lambda$. We will see in Sec.~\ref{sec_comparison} that this composition law corresponds in fact to $\kappa$-Poincaré.

When $\epsilon=-1$, one has
\be
\left(1-\frac{(p^{(1)}\oplus p^{(2)})_0}{\Lambda}\right) \,=\, \left(1-\frac{p^{(1)}_0}{\Lambda}\right) \left(1-\frac{p^{(2)}_0}{\Lambda}\right)\,,
\ee
so that the scale $\Lambda$ plays the role of a cutoff in the energy. This is the reason why this choice of sign reproduces the DCL in the context of DSR, as we will later see.  The other choice of sign $\epsilon=+1$ corresponds to a deformation where the scale $\Lambda$ is not a maximum of the energy and it goes beyond the framework of DSR.


If we go back to the expression for the relative generalized space-time coordinates \eqref{xtilde} and use the explicit form of the local DCL1 in Eq.~\eqref{DCL(1)}, we find
\begin{align}
  & \lim_{k\to 0} \frac{\partial(p^{(1)}\oplus k)_0}{\partial k_0} \,=\, 1 + \epsilon \frac{p^{(1)}_0}{\Lambda}\,, \quad \lim_{k\to 0} \frac{\partial(p^{(1)}\oplus k)_0}{\partial k_i} \,=\, \lim_{k\to 0} \frac{\partial(p^{(1)}\oplus k)_i}{\partial k_0} \,=\, 0\,, \quad \lim_{k\to 0} \frac{\partial(p^{(1)}\oplus k)_i}{\partial k_j} \,=\, \delta_i^j \left(1 + \epsilon \frac{p^{(1)}_0}{\Lambda}\right)\,, \nonumber \\
  & \lim_{k\to 0} \frac{\partial(k\oplus p^{(2)})_0}{\partial k_0} \,=\, 1 + \epsilon \frac{p^{(2)}_0}{\Lambda}\,, \quad \lim_{k\to 0} \frac{\partial(k\oplus p^{(2)})_0}{\partial k_i} \,=\, 0\,, \quad \lim_{k\to 0} \frac{\partial(k\oplus p^{(2)})_i}{\partial k_0} \,=\, \epsilon \frac{p^{(2)}_i}{\Lambda}\,, \quad \lim_{k\to 0} \frac{\partial(k\oplus p^{(2)})_i}{\partial k_j} \,=\, \delta^j_i\,,
\end{align}
and then
\be
\tilde{x}^0_{(12)} \,=\, x_{(1)}^0 (1+\epsilon p^{(1)}_0/\Lambda) - x_{(2)}^0 (1+\epsilon p^{(2)}_0/\Lambda) - x_{(2)}^j \epsilon p^{(2)}_j/\Lambda\,, \quad\quad
\tilde{x}^i_{(12)} \,=\, x_{(1)}^i (1+\epsilon p^{(1)}_0/\Lambda) - x_{(2)}^i\,.
\ee
From these expressions for the relative space-time coordinates, we have
\begin{align}
\{\tilde{x}^i_{(12)}, \tilde{x}^0_{(12)}\} \,=& \{x_{(1)}^i (1+\epsilon p^{(1)}_0/\Lambda), x_{(1)}^0 (1+\epsilon p^{(1)}_0/\Lambda)\} + \{x_{(2)}^i, x_{(2)}^j \epsilon p^{(2)}_j/\Lambda\} \nonumber \\ =& (\epsilon/\Lambda) \,\left[x_{(1)}^i (1+\epsilon p^{(1)}_0/\Lambda) - x_{(2)}^i\right] \,=\, (\epsilon/\Lambda) \,\tilde{x}^i_{(12)}\,.
\end{align}
Then we see that the relative space-time coordinates of the two-particle system are the coordinates of a (noncommutative) $\kappa$-Minkowski spacetime with $\kappa=\epsilon/\Lambda$. 

If we want to determine the generalized space-time coordinates of the two-particle system (not just the relative coordinates), we have to solve, using the explicit form of the local DCL1, the system of equations in (\ref{loc-oplus-varphi}) for the functions $\varphi^\alpha_\mu(p)$ which define the generalized space-time coordinates of a one-particle system. One has different solutions and then different choices for generalized space-time coordinates with a crossing of worldlines. In order to have a well-defined spacetime defined by locality, one has to include an additional requirement.

The expression of the deformed composition law in (\ref{DCL(1)})
\be
\left(p^{(1)}\oplus p^{(2)}\right)_\mu \,=\, p^{(1)}_{\:\mu} + \left(1 + \epsilon p^{(1)}_0/\Lambda\right) \,p^{(2)}_{\:\mu}\,,
\ee
is a sum of $p_{\:\mu}^{(1)}$ (independent of $p^{(2)}$) and a term proportional to $p_{\:\mu}^{(2)}$ depending on $p^{(1)}$. This suggests to consider generalized space-time coordinates $\tilde{x}^{\:\mu}_{(1)}$ depending on the phase-space coordinates ($x_{(1)}, p^{(1)}$), while $\tilde{x}^{\:\mu}_{(2)}$ depends on the phase-space coordinates of both particles ($x_{(1)}, p^{(1)}, x_{(2)}, p^{(2)}$), as the additional requirement to derive the generalized space-time coordinates of the two-particle system. In this case one has
\be
\varphi^{(2)\alpha}_{(1)\mu}(p^{(2)}) \,=\, 0\,, \quad\quad \rightarrow \quad\quad \varphi^\alpha_\mu(p^{(1)}) \,=\, \lim_{k\to 0} \frac{\partial(k\oplus p^{(1)})_\mu}{\partial k_\alpha}\,,
\label{simplphi}
\ee
and
\begin{align}
\varphi^\alpha_\mu(p^{(1)}\oplus p^{(2)}) \,=& \lim_{k\to 0} \frac{\partial(k\oplus(p^{(1)}\oplus p^{(2)}))_\mu}{\partial k_\alpha} \,=\, \lim_{k\to 0} \frac{\partial((k\oplus p^{(1)})\oplus p^{(2)})_\mu}{\partial k_\alpha} \,=\, \lim_{k\to 0} \left[\frac{\partial((k\oplus p^{(1)})\oplus p^{(2)})_\mu}{\partial (k\oplus p^{(1)})_\nu} \frac{\partial (k\oplus p^{(1)})_\nu}{\partial k_\alpha}\right] \nonumber \\ =& \frac{\partial(p^{(1)}\oplus p^{(2)})_\mu}{\partial p^{(1)}_\nu} \lim_{k\to 0} \frac{\partial (k\oplus p^{(1)})_\nu}{\partial k_\alpha}\,,
\label{chainrule}
\end{align}
where we have made use of the associativity of the local DCL1. But then one has [using \eqref{simplphi} and \eqref{chainrule}]:
\be
\frac{\partial(p^{(1)}\oplus p^{(2)})_\mu}{\partial p^{(1)}_\nu} \varphi^\alpha_\nu(p^{(1)}) + 
\frac{\partial(p^{(1)}\oplus p^{(2)})_\mu}{\partial p^{(2)}_\nu} \varphi^\alpha_\nu(p^{(2)}) - \varphi^\alpha_\mu(p^{(1)}\oplus p^{(2)}) \,=\, 
\frac{\partial(p^{(1)}\oplus p^{(2)})_\mu}{\partial p^{(2)}_\nu} 
\lim_{k\to 0} \frac{\partial(k\oplus p^{(2)})_\nu}{\partial k_\alpha}\,,
\ee
which is the relation between $\varphi^\alpha_\mu$ and the DCL ($\oplus$) which results from the requirement to have a crossing of worldlines, Eq.~\eqref{loc-oplus-varphi}. Then we have proved that in the case of DCL1 it is possible to implement the locality of interactions through a mixing of the phase spaces of the two particles in the spacetime of just one of the particles. The functions defining the generalized space-time coordinates for a one-particle system are
\be
\varphi^\alpha_\mu(p) \,=\, \lim_{k\to 0} \frac{\partial(k\oplus p)_\mu}{\partial k_\alpha}\,,
\label{magicformula}
\ee
and then, using the composition of four-momenta in (\ref{DCL(1)}), we have
\begin{align}
\tilde{x}^0 \,=& x^\mu \lim_{k\to 0} \frac{\partial(k\oplus p)_\mu}{\partial k_0} \,=\, x^0 (1 + \epsilon p_0/\Lambda) + x^j \epsilon p_j/\Lambda\,, \nonumber \\
\tilde{x}^i \,=& x^\mu \lim_{k\to 0} \frac{\partial(k\oplus p)_\mu}{\partial k_i} \,=\, x^i\,,
\label{eq:tilde_1}
\end{align}
and
\be
\{\tilde{x}^i, \tilde{x}^0\} \,=\, \{x^i, x^j \epsilon p_j/\Lambda\} \,=\, - (\epsilon/\Lambda) x^i \,=\, - (\epsilon/\Lambda) \tilde{x}^i\,.
\label{xtilde-}
\ee
The space-time coordinates of a one-particle system are also the coordinates of a (noncommutative) $\kappa$-Minkowski spacetime with $\kappa=-(\epsilon/\Lambda)$.

If one considers the second case for a deformed composition law quadratic in momenta and compatible with the implementation of locality,
  \be
  \left(p^{(1)}\oplus p^{(2)}\right)_\mu \,=\, \left(1 + \epsilon p^{(2)}_0/\Lambda\right) \,p^{(1)}_{\:\mu} + p^{(2)}_{\:\mu}\,,
\label{DCL(1')}
  \ee
  one has now a sum of $p^{(2)}_{\:\mu}$ (independent of $p^{(1)}$) and a term proportional to $p^{(1)}_{\:\mu}$ depending on $p^{(2)}$. Then, one can consider generalized space-time coordinates $\tilde{x}^{\:\mu}_{(2)}$ depending on the phase-space coordinates ($x_{(2)}, p^{(2)}$), while $\tilde{x}^{\:\mu}_{(1)}$ depends on the phase-space coordinates of both particles ($x_{(1)}, p^{(1)}, x_{(2)}, p^{(2)}$) as the additional requirement. In this case one has 
\be
\varphi^{(1)\alpha}_{(2)\mu}(p^{(1)}) \,=\, 0\,, \quad\quad \rightarrow \quad\quad \varphi^\alpha_\mu(p^{(1)}) \,=\, \lim_{k\to 0} \frac{\partial(p^{(1)}\oplus k)_\mu}{\partial k_\alpha}\,.
\ee
The functions defining the generalized space-time coordinates for a one-particle system are
\be
\varphi^\alpha_\mu(p) \,=\, \lim_{k\to 0} \frac{\partial(p\oplus k)_\mu}{\partial k_\alpha}\,,
\ee
and then, using the composition of four-momenta in (\ref{DCL(1')}), we have
\begin{align}
\tilde{x}^0 \,=& x^\mu \lim_{k\to 0} \frac{\partial(p\oplus k)_\mu}{\partial k_0} \,=\, x^0 (1 + \epsilon p_0/\Lambda) + x^j \epsilon p_j/\Lambda\,, \nonumber \\
\tilde{x}^i \,=& x^\mu \lim_{k\to 0} \frac{\partial(p\oplus k)_\mu}{\partial k_i} \,=\, x^i\,.
\label{eq:tilde_2}
\end{align}
The expressions of the generalized space-time coordinates of the one-particle system in terms of the canonical phase-space coordinates are the same in the two cases.

\section{Local DCL1 as a relativistic kinematics}
\label{sec_rel_kinematics}


Until now, we have discussed one of the ingredients in a deformation of SR kinematics, the modification of the composition law for the four-momentum and its relation with the locality of interactions. We now consider the compatibility of the conservation of the total four-momentum in an interaction with Lorentz invariance. We have to consider a non-linear implementation of Lorentz transformations in the two-particle system, which will be defined by the expression of the six generators $J^{\alpha\beta}$ as functions of the two-particle phase-space coordinates 
\be
J^{\alpha\beta} \,=\, x^\mu_{(1)} {\cal J}^{\alpha\beta}_{(1)\mu}(p^{(1)}, p^{(2)}) + 
x^\mu_{(2)} {\cal J}^{\alpha\beta}_{(2)\mu}(p^{(1)}, p^{(2)})\,.
\ee
The action of Lorentz transformations on the two-particle system is given by
\begin{align}
& \{p^{(1)}_\mu, J^{\alpha\beta}\} \,=\, {\cal J}^{\alpha\beta}_{(1)\mu}(p^{(1)}, p^{(2)})\,, \quad\quad \{x_{(1)}^\mu, J^{\alpha\beta}\} \,=\, - x^\nu_{(1)} \frac{\partial {\cal J}^{\alpha\beta}_{(1)\nu}(p^{(1)}, p^{(2)})}{\partial p^{(1)}_\mu} - x^\nu_{(2)} \frac{\partial {\cal J}^{\alpha\beta}_{(2)\nu}(p^{(1)}, p^{(2)})}{\partial p^{(1)}_\mu}\,, \\ \nonumber
  & \{p^{(2)}_\mu, J^{\alpha\beta}\} \,=\, {\cal J}^{\alpha\beta}_{(2)\mu}(p^{(1)}, p^{(2)})\,, \quad\quad \{x_{(2)}^\mu, J^{\alpha\beta}\} \,=\, - x^\nu_{(1)} \frac{\partial {\cal J}^{\alpha\beta}_{(1)\nu}(p^{(1)}, p^{(2)})}{\partial p^{(2)}_\mu} - x^\nu_{(2)} \frac{\partial {\cal J}^{\alpha\beta}_{(2)\nu}(p^{(1)}, p^{(2)})}{\partial p^{(2)}_\mu}\,.
\end{align}
In the one-particle system, the generators of Lorentz transformations will be given in terms of the phase-space coordinates by
\be
J^{\alpha\beta} \,=\, x^\mu {\cal J}^{\alpha\beta}_\mu(p)\,,
\ee
and one has
\be
\{p_\mu, J^{\alpha\beta}\} \,=\, {\cal J}^{\alpha\beta}_\mu(p)\,, \quad\quad \{x^\mu, J^{\alpha\beta}\} \,=\, - x^\nu \frac{\partial{\cal J}^{\alpha\beta}_\nu(p)}{\partial p_\mu}\,.
\ee
The identification of the one-particle system with a two-particle system when one of the four-momenta is zero leads to the relations
\begin{align}
& {\cal J}^{\alpha\beta}_{(1)\mu}(p^{(1)}, 0) \,=\, {\cal J}^{\alpha\beta}_\mu(p^{(1)})\,, \quad\quad {\cal J}^{\alpha\beta}_{(1)\mu}(0, p^{(2)}) \,=\, 0\,, \nonumber \\
& {\cal J}^{\alpha\beta}_{(2)\mu}(p^{(1)}, 0) \,=\, 0\,, \quad\quad {\cal J}^{\alpha\beta}_{(2)\mu}(0, p^{(2)}) \,=\, {\cal J}^{\alpha\beta}_\mu(p^{(2)})\,.
\end{align}
The compatibility of the conservation of the total four-momentum with Lorentz invariance requires that
\be
\{(p^{(1)}\oplus p^{(2)})_\mu, J^{\alpha\beta}\} \,=\, \frac{\partial(p^{(1)}\oplus p^{(2)})_\mu}{\partial p^{(1)}_\nu} \{p^{(1)}_\nu, J^{\alpha\beta}\} + \frac{\partial(p^{(1)}\oplus p^{(2)})_\mu}{\partial p^{(2)}_\nu} \{p^{(2)}_\nu, J^{\alpha\beta}\}\,,
\ee    
where on the left-hand side one has the generators of Lorentz transformations in a one-particle system and on the right-hand side the generators in the two-particle system. Then, the conservation law for the four-momentum will be Lorentz invariant if one can find a solution to the system of equations
\be
{\cal J}^{\alpha\beta}_\mu(p^{(1)}\oplus p^{(2)}) \,=\, \frac{\partial(p^{(1)}\oplus p^{(2)})_\mu}{\partial p^{(1)}_\nu} \,  {\cal J}^{\alpha\beta}_{(1)\nu}(p^{(1)}, p^{(2)}) + 
\frac{\partial(p^{(1)}\oplus p^{(2)})_\mu}{\partial p^{(2)}_\nu} \, {\cal J}^{\alpha\beta}_{(2)\nu}(p^{(1)}, p^{(2)})\,,
\ee
for the functions of one or two four-momenta that define the nonlinear action of the Lorentz transformations on the four-momentum of a particle or on the four-momenta of a system of two particles.

In order to determine the Lorentz transformation of the two-particle system, one also needs an additional requirement as in the case of the generalized space-time coordinates. The identification of generalized space-time coordinates with a mixing of phase-space coordinates only on the coordinates of one of the particles ($\tilde{x}_{(2)}$) suggests to consider a Lorentz transformation where only the transformation of one of the four-momenta ($p^{(2)}$) depends on the four-momentum of both particles. One has in this case
\be
{\cal J}^{\alpha\beta}_{(1)\mu}(p^{(1)}, p^{(2)}) \,=\, {\cal J}^{\alpha\beta}_\mu(p^{(1)})\,,
\ee
and the system of equations for the Lorentz invariance of the conservation law becomes
\be
 \frac{\partial(p^{(1)}\oplus p^{(2)})_\mu}{\partial p^{(2)}_\nu} \, {\cal J}^{\alpha\beta}_{(2)\nu}(p^{(1)}, p^{(2)}) \,=\, {\cal J}^{\alpha\beta}_\mu(p^{(1)}\oplus p^{(2)}) - \frac{\partial(p^{(1)}\oplus p^{(2)})_\mu}{\partial p^{(1)}_\nu}  \, {\cal J}^{\alpha\beta}_\nu(p^{(1)})\,.
\label{LT-p(2)}
\ee
This is a system of equations allowing to determine, given the composition law for the four-momentum ($\oplus$), the Lorentz transformation of the two-particle system from the Lorentz transformation of a one-particle system.

One possibility to fix the Lorentz transformation of a one-particle system is to require that the Lorentz generators, together with the generalized space-time coordinates $\tilde{x}^\alpha$, generate a deformed ten-dimensional Lie algebra in correspondence with the Poincaré algebra generated by the space-time coordinates and the Lorentz generators in SR. One finds (see Appendix~\ref{LT-one-particle}):
\begin{align}
  & {\cal J}^{ij}_0(p) \,=\, 0\,, \quad\quad\quad {\cal J}^{ij}_k(p) \,=\, \delta^j_k \, p_i - \delta^i_k \, p_j\,, \nonumber \\
  & {\cal J}^{0j}_0(p) \,=\, - p_j (1+\epsilon p_0/\Lambda)\,, \quad\quad\quad {\cal J}^{0j}_k \,=\, \delta^j_k \left[-p_0 - \epsilon p_0^2/2\Lambda\right] + (\epsilon/\Lambda) \left[\vec{p}^2/2 - p_j p_k\right]\,.
\label{LT1}
\end{align}  
There is no effect of the deformation on the transformation under rotations as a consequence of the isotropy of the deformed composition law in (\ref{DCL(1)}).

From Eq.~(\ref{LT-p(2)}), and using the local DCL1 (\ref{DCL(1)}) and the Lorentz transformation of the one-particle system in (\ref{LT1}), we find for the Lorentz transformation of the particle with phase-space coordinates ($x_{(2)}^\mu, p^{(2)}_\mu$) in the two-particle system
\begin{align}
{\cal J}_{(2)0}^{\,0i}(p^{(1)}, p^{(2)})\,=&\left(1+ \epsilon p^{(1)}_0/\Lambda\right) {\cal J}_0^{0i}(p^{(2)})\,, \nonumber \\
 {\cal J}_{(2)j}^{\,0i}(p^{(1)}, p^{(2)})\,=&\left(1+\epsilon p^{(1)}_0/\Lambda\right) {\cal J}_j^{0i}(p^{(2)}) + (\epsilon/\Lambda) \,\left(p^{(1)}_j p^{(2)}_i - \delta^i_j \vec{p}^{(1)}\cdot\vec{p}^{(2)}\right)\,, \nonumber\\
{\cal J}_{(2)0}^{\,ij}(p^{(1)}, p^{(2)})\,=& {\cal J}^{ij}_0(p^{(2)})\,,  \quad\quad\quad
{\cal J}_{(2)k}^{\,ij}(p^{(1)}, p^{(2)})\,=\, {\cal J}^{ij}_k(p^{(2)})\,.
\label{calJ(2)}
\end{align}

Once we have determined a Lorentz transformation of the one- and two-particle systems compatible with the invariance of the conservation law of the total four-momentum, 
one can determine the DDR, defined by a function $C(p)$ which is Lorentz invariant, i.e., such that  
\be
\{C(p), J^{\alpha\beta}\} \,=\, \frac{\partial C(p)}{\partial p_\mu} \, {\cal J}^{\alpha\beta}_\mu(p) \,=\, 0\,.
\label{LI-DDR}
\ee 
The result (when one adds the requirement that in the limit $(p_0^2/\Lambda^2) \to 0$, $(\vec{p}^2/\Lambda^2)\to 0$ the function $C(p)$ reduces to $p_0^2 - \vec{p}^2$, so that one recovers the dispersion relation of SR in the low-energy limit) is
\be
C(p) \,=\, \frac{p_0^2 - \vec{p}^2}{(1 + \epsilon p_0/\Lambda)}\,.
\ee

It is nontrivial to check, using (\ref{calJ(2)}) and (\ref{LT1}), that 
\be
J^{\alpha\beta} \,=\, x^\mu_{(1)} \,{\cal J}^{\alpha\beta}_\mu(p^{(1)}) + x^\mu_{(2)} \,{\cal J}^{\alpha\beta}_{(2)\mu}(p^{(1)}, p^{(2)})\,,
\ee
is a representation of the Lorentz algebra and that 
\be
\frac{\partial C(p^{(2)})}{\partial p^{(2)}_\mu} \,{\cal J}^{\alpha\beta}_{(2)\mu}(p^{(1)}, p^{(2)}) \,=\, 0\,,
\ee
so that the Lorentz transformations which leave invariant the conservation of the total four-momentum leave also invariant the dispersion relations of the two particles. Then, we have shown that one has a relativistic deformed kinematics with a composition law DCL1, Eq.~(\ref{DCL(1)}), and a crossing of worldlines in the interaction of particles when one introduces generalized space-time coordinates $\tilde{x}^\alpha_{(1)}$ depending on the phase-space coordinates ($x_{(1)}^\mu, p^{(1)}_\mu$) and space-time coordinates $\tilde{x}^\alpha_{(2)}$ depending on all the phase space coordinates ($x_{(1)}^\mu, p^{(1)}_\mu, x^\mu_{(2)}, p^{(2)}_\mu$). The previous choice of generalized space-time coordinates can be combined with a Lorentz transformation of the momentum $p^{(1)}$ which does not depend on the second momentum $p^{(2)}$, while the Lorentz transformation of the momentum $p^{(2)}$ depends on both momenta. This is just an example of the different ways to implement locality and the relativity principle with the local DCL1 (\ref{DCL(1)}).

We end up this section by pointing out that the standard treatment of Lorentz invariance violation, based on a total four-momentum given by the sum of four-momenta (with a crossing of worldlines in the canonical spacetime) and a deformed dispersion relation with an additional energy scale $\Lambda$ (scale of Lorentz invariance violation) such that in the limit $(p_0^2/\Lambda^2)\to 0$, $(\vec{p}^2/\Lambda^2)\to 0$ one recovers the dispersion relation of SR, is just an example of deformed kinematics which is not compatible with the relativity principle. One can have other possibilities for Lorentz invariance violations with a total four-momentum differing from the sum of four-momenta but with a crossing of worldlines in a generalized spacetime if one considers a DDR which is not invariant under the Lorentz transformations determined by the DCL through the requirement to have a ten-dimensional Lie algebra with Lorentz generators and the generalized space-time coordinates as generators.     

\section{Local DCL1 and \texorpdfstring{$\kappa$}{k}-Poincaré kinematics}
\label{sec_comparison}

We have not considered, in all the discussion of the model with a deformed composition law ($\oplus$), the arbitrariness in the starting point corresponding to the choice of canonical coordinates in phase space. In fact, if one considers new momentum coordinates $p'_\mu$ related nonlinearly to $p_\nu$, then one will have a new dispersion relation defined by a function $C'$ and a new deformed composition law $\oplus'$ which are related to the function $C$ and the deformed composition law $\oplus$ by
\be
C(p) \,=\, C'(p')\,, \quad\quad\quad (p'^{(1)}\oplus' p'^{(2)})_\mu \,=\, (p^{(1)}\oplus p^{(2)})'_\mu\,.
\ee  
Then we have 
\be
\varphi'^\alpha_\mu(p') \,=\, \lim_{k'\to 0} \frac{\partial(k'\oplus' p')_\mu}{\partial k'_\alpha} \,=\, \lim_{k'\to 0} \frac{\partial(k\oplus p)'_\mu}{\partial k'_\alpha} \,=\, \lim_{k\to 0} \frac{\partial k_\beta}{\partial k'_\alpha} \frac{\partial(k\oplus p)'_\mu}{\partial k_\beta} \,=\, \lim_{k\to 0} \frac{\partial(k\oplus p)'_\mu}{\partial(k\oplus p)_\nu} \frac{\partial(k\oplus p)_\nu}{\partial k_\alpha} \,=\, \frac{\partial p'_\mu}{\partial p_\nu} \varphi^\alpha_\nu(p)\,,
\label{varphi'-varphi}
\ee
where we have used that $\partial k_\beta/\partial k'_\alpha=\delta_{\beta}^\alpha$ when $k\to 0$.
On the other hand, the nonlinear change of momentum variables $p\to p'$ defines a canonical change of coordinates in phase space with 
\be
x'^\mu \,=\, x^\rho \frac{\partial p_\rho}{\partial p'_\mu}\,,
\ee
and then 
\be
x'^\mu \varphi'^\alpha_\mu(p') \,=\, x^\rho \frac{\partial p_\rho}{\partial p'_\mu} \varphi'^\alpha_\mu(p') \,=\, x^\rho \frac{\partial p_\rho}{\partial p'_\mu} \frac{\partial p'_\mu}{\partial p_\nu} \varphi^\alpha_\nu(p) \,=\, x^\nu \varphi^\alpha_\nu(p)\,,
\ee
where we have used Eq.~\eqref{varphi'-varphi} in the second equality.
This means that $\tilde{x}'^\alpha=\tilde{x}^\alpha$, and the generalized space-time coordinates for a one-particle system are invariant under a canonical change of phase-space coordinates corresponding to a nonlinear change of momentum variables.

When one considers the two-particle system, one has
\be
\varphi'^{(2)\alpha}_{(1)\mu}(p'^{(2)}) \,=\, \varphi'^\alpha_\mu(p'^{(2)}) - \lim_{k'\to 0}  \frac{\partial(k'\oplus' p'^{(2)})_\mu}{\partial k'_\alpha}\,,
\ee
but the same argument used in (\ref{varphi'-varphi}) leads to identify the relation
\be
\lim_{k'\to 0}  \frac{\partial(k'\oplus' p'^{(2)})_\mu}{\partial k'_\alpha} \,=\, \frac{\partial p'^{(2)}_\mu}{\partial p^{(2)}_\nu} \,\lim_{k\to 0}  \frac{\partial(k\oplus p^{(2)})_\mu}{\partial k_\alpha}\,,
\ee
and then one has
\be
\varphi'^{(2)\alpha}_{(1)\mu}(p'^{(2)}) \,=\, \frac{\partial p'^{(2)}_\mu}{\partial p^{(2)}_\nu} \,\varphi^{(2)\alpha}_{(1)\nu}(p^{(2)})\,.
\ee
The two-particle canonical change of variables in the two-particle phase space,
\be
x'^\mu_{(1)} \,=\, x^\nu_{(1)} \,\frac{\partial p^{(1)}_\nu}{\partial p'^{(1)}_\mu}\,, \quad\quad
x'^\mu_{(2)} \,=\, x^\nu_{(2)} \,\frac{\partial p^{(2)}_\nu}{\partial p'^{(2)}_\mu}\,, 
\ee
will then leave invariant the generalized space-time coordinates of the two-particle system
\be
\tilde{x}'^\alpha_{(1)} \,=\, \tilde{x}^\alpha_{(1)}\,, \quad\quad\quad \tilde{x}'^\alpha_{(2)} \,=\, \tilde{x}^\alpha_{(2)}\,.
\ee
This means that all the results (crossing of worldlines, a $\kappa$-Minkowski noncommutative generalized spacetime, and a relativistic deformed kinematics) obtained in the previous sections for the local DCL1 (\ref{DCL(1)}) will apply to any other deformed composition law obtained from it by a nonlinear change of momentum variables. 


In particular, one can consider a nonlinear change of momentum variables $p_\mu \to p'_\mu$ such that 
\be
p_i \,=\, p'_i\,, \quad\quad\quad (1 + \epsilon p_0/\Lambda) \,=\, e^{\epsilon p'_0/\Lambda}\,.
\ee
The new composition law that results by applying this change of momentum variables to the local DCL1 is
\be
(p'^{(1)}\oplus' p'^{(2)})_0 \,=\, p'^{(1)}_0 + p'^{(2)}_0, \quad\quad\quad
(p'^{(1)}\oplus' p'^{(2)})_i \,=\, p'^{(1)}_i + e^{\epsilon p'^{(1)}_0/\Lambda} \,p'^{(2)}_0\,,
\label{DCL-bcb}
\ee
and one also has
\be
\frac{p_0^2 - \vec{p}^2}{(1+\epsilon p_0/\Lambda)} \,=\, \Lambda^2 \left(e^{\epsilon p'_0/\Lambda} + e^{-\epsilon p'_0/\Lambda} -2\right) - \vec{p}'^2 \,e^{-\epsilon p'_0/\Lambda}\,.
\label{C(p)-bcb}
\ee
The composition law of four-momenta $p'_\mu$ is (when $\epsilon=-1$) just the composition law corresponding to the coproduct of momentum in the bicrossproduct basis of the $\kappa$-Poincaré Hopf algebra~\cite{KowalskiGlikman:2002jr}, and the new function of momentum $C'(p')$ is just the function corresponding to the Casimir of the $\kappa$-Poincaré deformed algebra in such basis. Then, we see that a change of momentum variables, applied to the deformed relativistic kinematics obtained from an implementation of locality with a deformed composition law DCL1, leads to the $\kappa$-Poincaré kinematics. This establishes the relation between the algebraic framework based on Hopf algebras to go beyond SR kinematics and the framework presented in this work based on the possibility to identify a generalized spacetime where interactions are local.        


The new perspective of RDK based on locality leads, together with $\kappa$-Poincaré relativistic kinematics in the case $\epsilon=-1$, to a new option when $\epsilon=1$. The main new ingredient is that the new scale $\Lambda$ is not a cutoff on the energy in this case. This is a possibility which deserves to be considered and has been overlooked in the context of doubly special relativity.   


The relation between $\kappa$-Poincaré kinematics and locality was identified in a less direct way in a previous work~\cite{Carmona2018}. In that paper, locality is implemented by asking that the noncommutative coordinates do not mix space-time variables. When one adds the condition that the space-time coordinates of the one-particle system (obtained as a limit of the coordinates of the two-particle system when one of the momenta goes to zero) define a $\kappa$-Minkowski noncommutative spacetime, it is found that $\kappa$-Poincar\'e kinematics is compatible with the locality of interactions. In the present paper, the new implementation of locality, based on the use of generalized space-time coordinates defined as a sum of two terms, each one depending of the phase-space coordinates of each particle, has allowed us to derive the general form of a deformed composition law DCL1 compatible with locality, and $\kappa$-Minkowski as the spacetime defined by the relative coordinates of the two-particle system. 

A closely related result was the identification of a ``rigid'' translational symmetry~\cite{Kowalski-Glikman:2014wba} in an extension to $3+1$ dimensions of a $2+1$ dimensional model for a deformed relativistic kinematics. The identification of $\kappa$-Poincaré kinematics as a deformed relativistic kinematics obtained from a model with local interactions gives an explanation of these previous results. 


In a recent work~\cite{Carmona:2019fwf}, a derivation of an isotropic relativistic deformed kinematics (RDK) from the geometry of a maximally symmetric momentum space has been presented. This gives a complementary perspective to the derivation of a RDK from the locality of interactions in a generalized spacetime, as presented in this paper. The ten-dimensional Lie algebra with Lorentz generators and generalized space-time coordinates as generators, is just the Lie algebra of the generators of isometries in the four-dimensional maximally symmetric momentum space. The implementation of Lorentz transformations in the two-particle system with the requirement that the transformation of one of the four-momentum does not depend on the other four-momentum (a condition which was necessary to be able to derive the DRK and was justified by the choice of new space-time coordinates in the two-particle system with a mixing of phase space coordinates only for one of the particles) has a simple explanation in the geometric perspective. It is a consequence of the identification of the composition law with an isometry. The derivation of the relativistic invariance of the conservation of the total momentum is a consequence of the identification of a composition of isometries as an isometry~\cite{Carmona:2019fwf}. 

\section{Associativity of the composition law of momenta, locality and relativistic kinematics}
\label{sec:associativity}

We have shown in Sec.~\ref{sec_first_order} that associativity is a necessary condition for a DCL1 to be compatible with locality of interactions. A nonlinear change of momentum variables applied to an associative composition law preserves this property, so that any deformed composition law obtained from a local DCL1 by a nonlinear change of variables will also be associative. In Sec.~\ref{sec_st_locality} we have also shown that any associative DCL is compatible with locality. This raises the question whether associativity will be a property of any DCL compatible with locality. 

If one goes back to the first equality in (\ref{loc-oplus-varphi}) and takes a derivative with respect to $p^{(1)}_\rho$ on both sides, and one introduces the notation
\be
L^\alpha_\nu(p^{(2)}) \,\doteq\, \lim_{k\to 0} \frac{\partial(k\oplus p^{(2)})_\nu}{\partial k_\alpha}\,, \quad\quad\quad 
R^\alpha_\nu(p^{(1)}) \,\doteq\, \lim_{k\to 0} \frac{\partial(p^{(1)}\oplus k)_\nu}{\partial k_\alpha}\,,
\ee
one finds
\be
\frac{\partial^2(p^{(1)}\oplus p^{(2)})_\mu}{\partial p^{(2)}_\nu \partial p^{(1)}_\rho} \,L^\alpha_\nu(p^{(2)}) \,=\, 
\frac{\partial^2(p^{(1)}\oplus p^{(2)})_\mu}{\partial p_\nu^{(1)} \partial p_\rho^{(1)}} R^\alpha_\nu(p^{(1)}) + \frac{\partial(p^{(1)}\oplus p^{(2)})_\mu}{\partial p_\nu^{(1)}} \frac{\partial R_\nu^\alpha(p^{(1)})}{\partial p^{(1)}_\rho}\,. 
\ee
Next, one can take the limit $\lim_{p^{(1)}\to 0}$, leading to
\be
\frac{\partial L^\rho_\mu(p^{(2)})}{\partial p^{(2)}_\nu} \, L^\alpha_\nu(p^{(2)}) \,=\, \frac{L^{\alpha\rho}_\mu(p^{(2)})}{\Lambda} + L^\nu_\mu(p^{(2)}) \, \frac{c^{\rho\alpha}_\nu}{\Lambda}\,,
\label{L}
\ee
where
\be
\frac{L^{\alpha\rho}_\mu(p^{(2)})}{\Lambda} \,\doteq\, \lim_{p^{(1)}\to 0} \frac{\partial^{2}(p^{(1)}\oplus p^{(2)})_\mu}{\partial p_\alpha^{(1)} \partial p_\rho^{(1)}}\,,
\ee
denotes the coefficient of the term proportional to $p_\alpha^{(1)} p_\rho^{(1)}$ in $(p^{(1)}\oplus p^{(2)})_\mu$, and 
\be
\frac{c^{\rho\alpha}_\nu}{\Lambda} \,\doteq\, \lim_{p^{(1)}, p^{(2)} \to 0} \frac{\partial^2(p^{(1)}\oplus p^{(2)})_\nu}{\partial p^{(1)}_\rho \partial p^{(2)}_\alpha}\,,
\ee
is the coefficient of the term proportional to $p^{(1)}_\rho p^{(2)}_\alpha$ in $(p^{(1)}\oplus p^{(2)})_\nu$. Using the symmetry under the exchange $\alpha\leftrightarrow \rho$ of $L^{\alpha\rho}_\mu$ in (\ref{L}), one has
\be
\frac{\partial L^\rho_\mu(p^{(2)})}{\partial p^{(2)}_\nu} \, L^\alpha_\nu(p^{(2)}) - \frac{\partial L^\alpha_\mu(p^{(2)})}{\partial p^{(2)}_\nu} \, L^\rho_\nu(p^{(2)}) \,=\, \frac{(c^{\rho\alpha}_\nu - c^{\alpha\rho}_\nu)}{\Lambda} \, L^\nu_\mu(p^{(2)})\,.
\ee
This set of differential equations for the functions $L^\mu_\nu(p^{(2)})$ is just the condition that the phase-space variables 
\be 
T_L^\mu \,\doteq\, x^{(2)\rho} \,L^\mu_\rho(p^{(2)})\,,
\ee
are the generators of a Lie algebra 
\be
\{T_L^\mu, T_L^\nu \} \,=\, \frac{(c^{\mu\nu}_\rho - c^{\nu\mu}_\rho)}{\Lambda} \, T_L^\rho\,.
\ee
The infinitesimal transformation, with parameter $\epsilon$, of the momentum $p^{(2)}$ is
\be
\delta p^{(2)}_\mu \,=\, \epsilon_\nu \{p^{(2)}_\mu, T_L^\nu \} \,=\, \epsilon_\nu L^\nu_\mu(p^{(2)}) \,=\, \epsilon_\nu \lim_{k\to 0} \frac{\partial (k\oplus p^{(2)})_\mu}{\partial k_\nu} \,=\, (\epsilon\oplus p^{(2)})_\mu - p^{(2)}_\mu\,.
\ee

If the composition law is associative, then it can be used to define a finite transformation by successive application of the infinitesimal transformations generated by the $T_L^\mu$, as
\be
p^{(2)}_\mu \to p^{(2)'}_\mu \,=\, (a \oplus p^{(2)})_\mu\,,
\ee
for a transformation with parameter $a$.

Similarly, taking a derivative with respect to $p^{(2)}_\rho$ instead of $p^{(1)}_\rho$ in (\ref{loc-oplus-varphi}) and considering later the limit $\lim_{p^{(2)}\to 0}$, one finds that 
\be
T_R^\mu \doteq x^{(1)\nu} R^\mu_\nu(p^{(1)})
\ee
are the generators of a Lie algebra 
\be
\{T_R^\mu, T_R^\nu \} \,=\, - \,\frac{(c^{\mu\nu}_\rho - c^{\nu\mu}_\rho)}{\Lambda} \, T_R^\rho\,,
\ee
which is just the Lie algebra we have found for $T_L$ with a global sign changed in the structure constants. The infinitesimal transformation with parameter $\epsilon$ of the momentum $p^{(1)}$ is
\be
\delta p^{(1)}_\mu \doteq \epsilon_\nu \{p^{(1)}_\mu, T_R^\nu \} \,=\, \epsilon_\nu R^\nu_\mu(p^{(1)}) \,=\, \epsilon_\nu \,\lim_{k\to 0} \frac{\partial(p^{(1)}\oplus k)_\mu}{\partial k_\nu} \,=\, (p^{(1)}\oplus \epsilon)_\mu - p^{(1)}_\mu\,,
\ee
and if the composition law is associative, then it defines a finite transformation
\be
p^{(1)}_\mu \to p^{(1)'}_\mu \,=\, (p^{(1)}\oplus a)_\mu\,.
\ee

In particular, we have seen that a local DCL1, and the different composition laws obtained from it by a change of momentum variables, are associative, so that they can be associated to finite transformations generated by the $T_L^\mu$ or $T_R^\mu$. In Ref.~\cite{Carmona:2019fwf}, it was seen that $\kappa$-Poincaré kinematics corresponds to the only relativistic kinematics for which the generators of translations defined from the composition law close an algebra. This explains why the local DCL1 (which, as we saw in Section~\ref{sec_rel_kinematics}, defines a relativistic kinematics) is the deformed composition law of $\kappa$-Poincaré kinematics in a certain basis.

Other alternatives for a deformed relativistic kinematics beyond $\kappa$-Poincaré (Snyder and hybrid models) have been  obtained both in the algebraic~\cite{Meljanac:2009ej}  and geometric framework~\cite{Carmona:2019fwf}. They lead to $T_{L,R}^\mu$ generators (defined from the corresponding DCL) which do not close a Lie algebra and then, according to the previous arguments, are not compatible with the locality of interactions. Then we conclude that, in a deformed relativistic kinematics, the locality of interactions, implemented in the way proposed in this work, requires the associativity of the DCL and selects $\kappa$-Poincaré kinematics as the unique relativistic isotropic generalization of SR kinematics.

\section{Conclusions and prospects}
\label{sec:conclusions}


We have shown how a deformed composition law for the four-momentum defines, at the classical level, a modified notion of spacetime for a system of two particles through the crossing of worldlines in particle interactions, which is the way spacetime is introduced in SR.

We have used an appropriate nonlinear change of momentum variables to prove that the framework presented in this paper contains the results of the $\kappa$-Poincaré Hopf algebra and gives a complementary perspective to the derivation of a relativistic deformed kinematics based on the geometry of a maximally symmetric momentum space. 


The work presented in this paper opens a new perspective on the formulation of a deformation of relativistic quantum field theory based either on the implementation of a deformed kinematics with a deformed composition law for the four-momentum, or based on the associated generalized notion of spacetime.  


To have an understanding of the introduction of a new energy scale from different perspectives can be important in order to explore if it has a realization in Nature through its possible observable effects. The notion of spacetime based on the locality of interactions defined by a deformed composition for the four-momentum is just a reinterpretation of the classical model used to introduce the idea of relative locality (the loss of the locality of interaction in canonical spacetime due to a deformation of the composition of four-momenta). The identification of a physical spacetime which differs from the canonical spacetime has implications on the propagation of particles, in particular on the energy dependence of the velocity of propagation of a free particle~\cite{Carmona2018b,Carmona:2019oph}, which is one of the possible observable effects of a departure from SR with a new energy scale $\Lambda$.  

We have found some remaining ambiguities in the identification of the generalized two-particle space-time coordinates implementing the locality of interactions and also on the Lorentz transformations of the two-particle system. This suggests to look for some additional physical requirement replacing the adhoc prescription (mixing of phase space coordinates in the generalized space-time coordinates and mixing of the momentum variables in the Lorentz transformations for only one of the two particles) used in this work to fix those ambiguities.  

In contrast with the idea of the loss of the notion of absolute locality due to the modification of the energy-momentum conservation law in a deformed relativistic kinematics, we have proposed in this work the possibility to maintain the identification of a physical spacetime through the locality of interactions. The next step in this direction is to check the consistency of the new perspective of spacetime through the identification of observables related to the generalized space-time coordinates defined by the locality of interactions.  
 
\section*{Acknowledgments}
This work is supported by Spanish grants PGC2018-095328-B-I00 (FEDER/Agencia estatal de investigación), and DGIID-DGA No. 2015-E24/2.
The authors would also like to thank support from the COST Action CA18108, and acknowledge Giulia Gubitosi, Flavio Mercati and Giacomo Rosati for useful conversations.

\bibliography{QuGraPhenoBib}

\appendix

\section{Lorentz transformation in the one-particle system of the local DCL1 kinematics}
\label{LT-one-particle}

We define the Lorentz generators $J^{\alpha\beta}$ by the requirement that, together with the space-time coordinates $\tilde{x}^\alpha$, they generate a ten-dimensional Lie algebra. Given the Lorentz algebra generated by $J^{\alpha\beta}$ and the algebra of the space-time coordinates $\tilde{x}^\alpha$ (when there is no mixing of phase-space coordinates in $\tilde{x}_{(1)}^\alpha$) ,
\be
\{\tilde{x}^i, \tilde{x}^0\} \,=\, - (\epsilon/\Lambda) \,\tilde{x}^i\,, \quad\quad\quad \{\tilde{x}^i, \tilde{x}^j\} \,=\, 0\,,
\ee
the rest of the ten-dimensional algebra is determined by Jacobi identities:
\be
\{\tilde{x}^0, J^{0j}\} \,=\, \tilde{x}^j - (\epsilon/\Lambda) J^{0j}\,, \quad\quad \{\tilde{x}^i, J^{0j}\} \,=\, \delta^{ij} \tilde{x}^0 - (\epsilon/\Lambda) J^{ij}\,, \quad\quad \{\tilde{x}^0, J^{jk}\} \,=\, 0\,, \quad\quad \{\tilde{x}^i, J^{jk}\} \,=\, \delta^{ik} \tilde{x}^j - \delta^{ij} \tilde{x}^k\,.
\label{xtilde-J}
\ee
Using
\be
\{\tilde{x}^\alpha, J^{\beta\gamma}\} \,=\, \{x^\nu \varphi^\alpha_\nu(p), x^\rho {\cal J}^{\beta\gamma}_\rho(p)\} \,=\, x^\mu \left(\frac{\partial\varphi^\alpha_\mu(p)}{\partial p_\rho} {\cal J}^{\beta\gamma}_\rho(p) - \frac{\partial J^{\beta\gamma}_\mu(p)}{\partial p_\nu} \varphi^\alpha_\nu(p)\right)
\ee
in the algebra (\ref{xtilde-J}), one finds the set of equations 
\begin{align}
\left(\frac{\partial\varphi^0_\mu(p)}{\partial p_\rho} {\cal J}^{0j}_\rho(p) - \frac{\partial {\cal J}^{0j}_\mu(p)}{\partial p_\nu} \varphi^0_\nu(p)\right) \,=& \varphi^j_\mu(p) \blue{-} (\epsilon/\Lambda) {\cal J}^{0j}_\mu(p)\,, \nonumber \\
 \left(\frac{\partial\varphi^i_\mu(p)}{\partial p_\rho} {\cal J}^{0j}_\rho(p) - \frac{\partial {\cal J}^{0j}_\mu(p)}{\partial p_\nu} \varphi^i_\nu(p)\right) \,=& \delta^{ij} \varphi^0_\mu(p) \blue{-} (\epsilon/\Lambda) {\cal J}^{ij}_\mu(p)\,, \nonumber \\
\left(\frac{\partial\varphi^0_\mu(p)}{\partial p_\rho} {\cal J}^{jk}_\rho(p) - \frac{\partial {\cal J}^{jk}_\mu(p)}{\partial p_\nu} \varphi^0_\nu(p)\right) \,=& 0\,, \nonumber \\
\left(\frac{\partial\varphi^i_\mu(p)}{\partial p_\rho} {\cal J}^{jk}_\rho(p) - \frac{\partial {\cal J}^{jk}_\mu(p)}{\partial p_\nu} \varphi^i_\nu(p)\right) \,=& \delta^{ik} \varphi^j_\mu(p) - \delta^{ij} \varphi^k_\mu(p)\,.
\end{align}

Using the expressions
\begin{align}
& \varphi^0_0(p) \,=\, \lim_{k\to 0} \frac{\partial(k\oplus p)_0}{\partial k_0} \,=\, 1 + \epsilon p_0/\Lambda\,, \quad\quad  \varphi^j_0(p) \,=\, \lim_{k\to 0} \frac{\partial(k\oplus p)_0}{\partial k_j} \,=\, 0 \,,\nonumber \\
& \varphi^0_i(p) \,=\,   \lim_{k\to 0} \frac{\partial(k\oplus p)_i}{\partial k_0} \,=\, \epsilon p_i/\Lambda\,, \quad\quad \varphi^j_i(p) \,=\, \lim_{k\to 0} \frac{\partial(k\oplus p)_i}{\partial k_j} \,=\, \delta_i^j\,,
\end{align}
for the functions which define the generalized space-time coordinates in the one-particle system of the local DCL1 kinematics according to Eq.~\eqref{magicformula}, we end up with the following system of equations for ${\cal J}^{\alpha\beta}_\mu(p)$:
\begin{align}
  & \frac{\partial{\cal J}^{0j}_0}{\partial p_0} \,=\, (\epsilon/\Lambda) \,\left[2 {\cal J}^{0j}_0 - p_0 \frac{\partial{\cal J}^{0j}_0}{\partial p_0} - p_k \frac{\partial{\cal J}^{0j}_0}{\partial p_k}\right]\,, \quad\quad 
  \frac{\partial{\cal J}^{0j}_0}{\partial p_i} \,=\, - \delta^{ij} (1 + \epsilon p_0/\Lambda) + (\epsilon/\Lambda) {\cal J}^{ij}_0\,, \nonumber \\
  & \frac{\partial{\cal J}^{0j}_l}{\partial p_0} \,=\, - \delta^j_l + (\epsilon/\Lambda) \,\left[2 {\cal J}^{0j}_l - p_0 \frac{\partial{\cal J}^{0j}_l}{\partial p_0} - p_k \frac{\partial{\cal J}^{0j}_l}{\partial p_k}\right]\,, \quad\quad 
  \frac{\partial{\cal J}^{0j}_l}{\partial p_i} \,=\, - \delta^{ij} \epsilon p_l/\Lambda + (\epsilon/\Lambda) {\cal J}^{ij}_l\,, \nonumber \\
  & \frac{\partial{\cal J}^{jk}_0}{\partial p_0} \,=\, (\epsilon/\Lambda) \,\left[{\cal J}^{jk}_0 - p_0 \frac{\partial{\cal J}^{jk}_0}{\partial p_0} - p_m \frac{\partial{\cal J}^{jk}_0}{\partial p_m}\right]\,, \quad\quad 
  \frac{\partial{\cal J}^{jk}_0}{\partial p_i} \,=\, 0\,, \nonumber \\
& \frac{\partial{\cal J}^{jk}_l}{\partial p_0} \,=\, (\epsilon/\Lambda) \,\left[{\cal J}^{jk}_l - p_0 \frac{\partial{\cal J}^{jk}_l}{\partial p_0} - p_m \frac{\partial{\cal J}^{jk}_l}{\partial p_m}\right]\,, \quad\quad 
  \frac{\partial{\cal J}^{jk}_l}{\partial p_i} \,=\, \delta^{ij} \delta^k_l - \delta^{ik} \delta^j_l\,.
  \end{align}

When one adds the condition that in the limit $(p_0^2/\Lambda^2)\to 0$, $(\vec{p}^2/\Lambda^2)\to 0$, one should recover the SR Lorentz transformation
\be
   {\cal J}^{0j}_0 \to - p_j\,, \quad\quad {\cal J}^{0j}_k \to - \delta^j_k \,p_0\,, \quad\quad {\cal J}^{jk}_0 \to 0\,, \quad\quad {\cal J}^{jk}_l \to (\delta^k_l p_j - \delta^j_l p_k)\,,
   \ee
one finds a unique solution for the Lorentz transformation of the relativistic kinematics defined by the local DCL1:
\begin{align}
  & {\cal J}^{ij}_0(p) \,=\, 0\,, \quad\quad\quad {\cal J}^{ij}_k(p) \,=\, \delta^j_k \, p_i - \delta^i_k \, p_j\,, \nonumber \\
  & {\cal J}^{0j}_0(p) \,=\, - p_j (1+\epsilon p_0/\Lambda)\,, \quad\quad\quad {\cal J}^{0j}_k \,=\, \delta^j_k \left[-p_0 - \epsilon p_0^2/2\Lambda\right] + (\epsilon/\Lambda) \left[\vec{p}^2/2 - p_j p_k\right]\,.
\end{align}

\end{document}